\documentclass[preprint,prd,aps,nofootinbib,amssymb]{revtex4}
\usepackage{graphicx}
\usepackage{epsfig,amssymb,amsmath}
\newcommand{\beq}{\begin{equation}}
\newcommand{\eeq}{\end{equation}}
\newcommand{\bea}{\begin{eqnarray}}
\newcommand{\eea}{\end{eqnarray}}
\newcommand{\bear}{\begin{array}}
\newcommand {\eear}{\end{array}}
\newcommand{\bef}{\begin{figure}}
\newcommand {\eef}{\end{figure}}
\newcommand{\bec}{\begin{center}}
\newcommand {\eec}{\end{center}}

\def\EQ#1{Eq.~(\ref{#1})}

\def\REF#1{(\ref{#1})}
\def\GEV#1{10^{#1}{\rm\,GeV}}

\def\lrf#1#2{ \left(\frac{#1}{#2}\right)}
\def\lrfp#1#2#3{ \left(\frac{#1}{#2} \right)^{#3}}

\def\dd#1#2{\frac{\partial #1}{\partial #2}}

\def\dd#1#2{\frac{\partial #1}{\partial #2}}

\begin{document}
\draft
\tighten
\preprint{
DESY 14-022, TU-955, IPMU14-0044
}
\title{\large \bf
The 7 keV axion dark matter and the X-ray line signal
}
\author{
    Tetsutaro Higaki\,$^a$\footnote{email: thigaki@post.kek.jp},
    Kwang Sik Jeong\,$^b$\footnote{email: kwangsik.jeong@desy.de},
    Fuminobu Takahashi\,$^{c,d}$\footnote{email: fumi@tuhep.phys.tohoku.ac.jp}
    }
\affiliation{
 $^a$ Theory Center, KEK, 1-1 Oho, Tsukuba, Ibaraki 305-0801, Japan \\
 $^b$ Deutsches Elektronen Synchrotron DESY, Notkestrasse 85,
         22607 Hamburg, Germany \\
 $^c$ Department of Physics, Tohoku University, Sendai 980-8578, Japan\\
 $^d$ Kavli IPMU, TODIAS, University of Tokyo, Kashiwa 277-8583, Japan
    }

\vspace{2cm}

\begin{abstract}
We propose a scenario where the saxion dominates the energy density of the Universe and reheats
the standard model sector via the dilatonic coupling, while its axionic partner contributes to dark
matter decaying into photons via the same operator in supersymmetry. 
Interestingly, for the axion mass $m_a \simeq 7$\,keV and the decay constant $f_a \simeq 10^{14-15}$\,GeV,
the recently discovered X-ray line at $3.5$\,keV in the XMM Newton X-ray observatory data
can be explained. We discuss various  cosmological aspects of the $7$\,keV axion dark matter such as
the production of axion dark matter, 
the saxion decay process, hot dark matter and isocurvature constraints on the axion
dark matter, and the possible baryogenesis scenarios. 
\end{abstract}

\maketitle

\section{Introduction}

In supergravity and superstring theories there appear many moduli fields at low energy scale
through compactifications of extra dimensions~\cite{fluxc}.
Moduli fields must be stabilized to obtain a sensible low-energy theory,
and it is known that many of them are fixed by flux compactifications and acquire a heavy
mass~\cite{Grana:2005jc}.
The remaining light moduli not fixed by the fluxes can be stabilized either by instantons/gaugino
condensations {\it a la}
KKLT~\cite{Kachru:2003aw} or by supersymmetry (SUSY) breaking
effects~\cite{Berg:2005yu,Balasubramanian:2005zx,Conlon:2006tq,Choi:2006za}.\footnote{
For instance the QCD axion could be the axion component
of such a modulus field mainly stabilized by the SUSY breaking effects.}
Such  light moduli fields  may play an important role in  cosmology; some of them may dominate 
the Universe and decay into the standard model (SM) sector, or others could contribute to dark matter
or dark energy if their masses are sufficiently light.

Recently an unidentified X-ray line at about $3.5$~keV in the XMM-Newton X-ray observatory data of
various galaxy clusters and the Andromeda galaxy was reported independently by two
groups~\cite{Bulbul:2014sua,Boyarsky:2014jta}.
While there are a variety of systematic uncertainties that can affect the observed line energy and
flux, it is intriguing that the X-ray line can be explained by decaying dark matter such as
sterile neutrinos\footnote{
Recently, Ishida and two of the present authors (KSJ and FT) showed that the small mass and mixing 
of sterile neutrino dark matter suggested by the X-ray line can be easily realized by the split 
flavor mechanism where the breaking of flavor symmetry is tied to the breaking of the $B-L$ 
symmetry~\cite{Ishida:2014dlp}.
}~\cite{Dolgov:2000ew} or moduli fields~\cite{Hashiba:1997rp,Asaka:1997rv,Kasuya:2001tp,Kusenko:2012ch}.

The observations suggest the mass and the lifetime of the dark matter 
as~\cite{Bulbul:2014sua,Boyarsky:2014jta}:
\bea
m_{\rm DM} &\simeq&7 {\rm \, keV},\\
\tau_{\rm DM} &\simeq& 2 \times 10^{27}  - 2 \times 10^{28}\, {\rm sec},
\label{tau-obs}
\eea
where we have used the values obtained by the M31 data~\cite{Boyarsky:2014jta}, and
we adopt them as reference values in the following analysis assuming that decaying dark matter  
is the origin of the $3.5$\,keV X-ray line.

The light dark matter mass about $7$\,keV may be due to some approximate symmetry forbidding the mass.
We focus on the axion component of a modulus field $\Phi = (\sigma + i a)/\sqrt{2}$ stabilized 
by SUSY breaking effects, where $\sigma$ and $a$ are the saxion and the axion components, respectively.
The axion $a$ can remain extremely light as a result of the axionic shift symmetry,
\beq
\Phi \rightarrow \Phi +  i C,
\label{shiftsym}
\eeq
where $C$ is a real transformation parameter. The axion can acquire a small but non-zero mass of $7$\,keV
from some non-perturbative effects which explicitly break the above shift symmetry. 
We shall see that, if the modulus field $\Phi$ is coupled to the SM gauge fields with a decay constant
of order $\GEV{14-15}$, the lifetime of the axion falls in the  range of (\ref{tau-obs}), explaining
 the observed X-ray line.
On the other hand, the saxion $\sigma$ generically acquires a mass of order of the gravitino mass from 
SUSY breaking effects. The gravitino mass is not known, but it must be heaver than the electroweak 
scale in the gravity or anomaly mediation.
We assume this is the case throughout this letter.

The mass hierarchy  between the saxion and the axion leads to a unified picture of the cosmological
role of light moduli fields:  the saxion dominates the Universe and reheats the SM sector via
the dilatonic coupling, while the axion contributes to dark matter decaying into photons via the same operator in SUSY.
As we shall see shortly, the right abundance of axion dark matter can be produced by coherent oscillations
for the saxion mass about $\GEV{6}$ and the decay constant $f_a \simeq \GEV{14-15}$ without fine-tuning of the
initial misalignment angle.  We shall also see that the axions are generically produced by the saxion decay, 
which may contribute to hot dark matter (HDM) component in agreement with the recent 
observations~\cite{Wyman:2013lza,Hamann:2013iba, Battye:2013xqa}.
Therefore, the detailed study of the decaying axion dark matter via the X-ray observation and 
the observations of large-scale structure can be a probe of not only the nature of dark matter but also 
the reheating of the Universe as well as the high-energy physics close to the GUT scale.

In this letter we propose a scenario in which the $7$\,keV axion dark matter decaying into photons 
explains the origin of the $3.5$\,keV X-ray line, while the saxion dominates the Universe 
and reheats the SM sector via the same dilatonic coupling in SUSY.
We will study various aspects of this scenario, focusing on the saxion cosmology, the production 
mechanism of the axion dark matter,  the isocurvature  and HDM constraints, and possible baryogenesis scenarios in turn.

Lastly let us briefly mention the differences of our work from Ref.~\cite{Kusenko:2012ch}  (and other
works~\cite{Hashiba:1997rp,Asaka:1997rv,Kasuya:2001tp}). One of the main differences is the
SUSY breaking scale, i.e., the gravitino mass.  They focused on the light gravitino mass between keV and MeV, and
consider moduli dark matter with a similar mass,
which corresponds to the real component of the moduli, i.e.  the saxion, in our scenario.
On the other hand, it is its axionic partner that becomes dark matter in our scenario.
As long as the implications for the observation of the X-ray line are concerned, there is no significant difference between
these two models.   The crucial difference
 is that the heavy gravitino we consider enables a scenario in which the saxion dominates and reheats the Universe via the
 same dilatonic coupling in SUSY. Then we can unambiguously discuss the saxion and axion cosmology.

\section{Moduli stabilization and  Light Axion}
\label{sec:2}

We consider KKLT-type flux compactifications on a Calabi-Yau space~\cite{Kachru:2003aw}
where the dilaton and complex structure
moduli are stabilized by closed string fluxes.
The low energy effective theory of complexified K\"ahler moduli $X_I$ possesses perturbative shift symmetries,
and is described by the K\"ahler potential of no-scale form
 at the leading order of string coupling and $\alpha'$-corrections:
\bea
K = -2\ln {\cal V}_{\rm CY}(X_I+X^*_I),
\eea
where the Calabi-Yau volume ${\cal V}_{\rm CY}$ is a homogeneous function of
degree $3/2$ in $X_I+X^*_I$.
The shift symmetry makes ${\rm Im}(X_I)$ massless until non-perturbative effects
are added.
To have a light string axion, we clearly need some mechanism to stabilize its scalar partner, 
the saxion, while preserving the associated shift symmetry.

An interesting possibility is to stabilize the saxion by K\"ahler potential
in the presence of sequestered uplifting sector~\cite{Conlon:2006tq,Choi:2006za}.
This works when the superpotential includes non-perturbative terms to stabilize K\"ahler
moduli as in the original KKLT, but with smaller number of terms than the number of
K\"ahler moduli.
Let us consider the case where there are $n-1$ non-perturbative superpotential terms
for $n$ K\"ahler moduli.
Then appropriate field redefinition leads to
\bea
K &=& K(\Phi+\Phi^*,X_i+X^*_i),
\nonumber \\
W &=& \omega_0 + \sum_i A_i e^{-a_i X_i},
\label{Wxi}
\eea
for $X_I=(\Phi,X_i)$,
where we have included a constant superpotential, $\omega_0$, which
is originated from background fluxes.
For string compactification allowing 
\bea
\partial_\Phi K &=& 0,\\
\partial_{X_i} W + (\partial_{X_i} K)W &=& 0,
\eea
there exists a supersymmetric field configuration, and consequently all the K\"ahler moduli are
stabilized at a dS vacuum with a vanishingly small cosmological constant
after adding sequestered uplifting potential,
\bea
V_{\rm up} = \epsilon\, e^{2K/3},
\eea
where $\epsilon = {\cal O}(\omega_0^2)$ is chosen to cancel the cosmological constant.
The K\"ahler moduli $X_i$ acquire large supersymmetric masses around $\ln(M_p/m_{3/2}) \times m_{3/2}$
from the non-perturbative superpotential terms,
where $m_{3/2}=\langle e^{K/2}W \rangle$ is the gravitino mass and $M_p$ denotes the reduced
Planck scale.\footnote{
We take the Planck scale to be unity unless otherwise stated.
}
On the other hand, $\Phi$ is fixed by the condition $\partial_\Phi K=0$.
The saxion is relatively light compared to $X_i$, and the axion remains
massless due to the shift symmetry:
\bea
\label{ms}
m_\sigma &\simeq& \sqrt2 m_{3/2},
\nonumber \\
m_a &=& 0,
\eea
for $\Phi=\langle \Phi \rangle + (\sigma+i a)/\sqrt2$.
The fermionic component has mass approximately equal to $m_{3/2}$.
It is important to note that these results follow from the no-scale structure,
and are insensitive to the precise form of the K\"ahler potential \cite{Choi:2006za}.
One may consider more general K\"ahler potential, for which the saxion is
stabilized in a similar way, and its mass is of order of the gravitino mass~\cite{Higaki:2011me}.

To make the axion massive, one can introduce small non-perturbative effects involving
$\Phi$ so that the associated shift symmetry is explicitly broken:
\bea
\Delta W = A e^{-\sum_i b_i X_i} e^{-b \Phi},
\eea
for real constants $b$ and $b_i$. 
If the dynamical scale is below the gravitino mass, we need to consider the non-perturbative
dynamics in a non-SUSY framework.
In the following we will simply assume that the axion acquires a small mass, $m_a\simeq 7$\,keV,
as a result of  some non-perturbative dynamics.
For instance, it can be induced by hidden gauge interactions to which $\Phi$ is coupled.
Note that the axion cannot be the QCD axion because of its mass.
The large mass hierarchy between the saxion and axion is achieved when
$\Delta W$ is much smaller than $m_{3/2}$ at the vacuum.

The axion dark matter of mass $7$\,keV should couple to photons in order to account
for the observed X-ray line.
The axion coupling to photons arises from the interaction
\bea
\label{FWW}
{\cal L} = \frac{1}{4} \int d^2\theta\, F(X_I) {\cal W}^{\alpha} {\cal W}_\alpha
+ {\rm h.c.},
\eea
where the gauge kinetic function linearly depends on the K\"ahler moduli:
\bea
F = k\Phi + \sum_i k_i X_i + {\rm constant},
\eea
as indicated by the perturbative shift symmetry.
Here $k$ and $k_i$ are real constants, and ${\cal W}_\alpha$ denotes the supersymmetric
field strength of the SM gauge fields.
The gauge kinetic functions for the SM gauge groups have been assumed to have 
the same dependence on the K\"ahler moduli, as would be required for the gauge coupling
unification.\footnote{
In general the gauge kinetic function can be different for each gauge group, which however
slightly weakens the relation between the axion dark matter decay and  the saxion decay
as there are more degrees of freedom. 
}
From the above interaction, one obtains the axion coupling to photons in the canonical basis,
\bea
{\cal L}_{\rm axion} = \frac{\alpha_{\rm EM}}{4\pi}
\frac{a}{f_a} F_{\mu\nu} \tilde F^{\mu\nu},
\eea
where the axion decay constant is determined by
\bea
f_a =
\frac{M_p}{4\sqrt2 \pi^2}\frac{1}{k}
\langle \partial_\Phi\partial_{\Phi^*} K \rangle^{1/2},
\label{fa}
\eea
where $F_{\mu\nu}$ is the electromagnetic field strength, and $\alpha_{\rm EM}$ is
its gauge coupling.
The value of $\partial_\Phi\partial_{\Phi^*} K$ depends on the details of the moduli stabilization,
especially on the volume of the Calabi-Yau space. 
In the current set-up, if there is a hidden gauge group with the rank of ${\cal O}(10)$ on the 
D-brane wrapping on the bulk cycle, it can be one order of magnitude smaller.  
Also $k$ can easily take a value larger or smaller than unity by a factor of $10$, if we allow some
mild tuning of the moduli fields $X_I$, as we have taken the field basis such that $X_i$ is the 
exponent of a non-perturbative superpotential term.
Therefore, the plausible range of $f_a$ is between $\GEV{14}$ and $\GEV{16}$.

The decay rate of the axion into photons is given by
\bea
\Gamma_{a\to \gamma\gamma} = \frac{\alpha^2_{\rm EM}}{64\pi^3}\frac{m^3_a}{f^2_a},
\eea
and therefore its lifetime is estimated to be
\bea
\tau_a \simeq 2\times 10^{28}\,{\rm sec}\times
\left(\frac{\alpha_{\rm EM}}{1/137}\right)^{-2}
\left(\frac{m_a}{7{\rm keV}}\right)^{-3}
\left(\frac{f_a}{5\times 10^{14}{\rm GeV}}\right)^2,
\eea
assuming that the axion mainly decays into photons via the above coupling.
Hence, the observed $3.5$\,keV X-ray line can be explained for $m_a\simeq 7$\,keV and
$f_a\simeq (2-5)\times 10^{14}$\,GeV which is within the expected range of (\ref{fa}).

\section{Cosmology of $7$ keV axion dark matter}

\subsection{Abundance of axion dark matter}

Let us discuss the production of the $7$\,keV axion dark matter. 
First let us estimate thermal production of axions. Applying the result for the QCD axion~\cite{Turner:1986tb,Masso:2002np,Graf:2010tv,Salvio:2013iaa}
to the $7$\,keV axion,
the axion abundance is
\bea
\Omega_a^{(th)} h^2 &\simeq&0.2 \lrf{\gamma_a}{10^{-2}} \lrf{106.75}{g_*} \lrf{m_a}{7{\rm \,keV}}
\lrfp{2 \times \GEV{14}}{f_a}{2} \lrf{T_R}{\GEV{12}},
\eea
where $\gamma_a$ is a numerical factor that parametrizes contributions from various sources, and
its typical value is between $0.01$  and $0.1$ for $\GEV{4} < T_R < \GEV{12}$~\cite{Salvio:2013iaa}.  $g_*$ counts the relativistic degrees of freedom at the reheating. 
As we shall see later, as long as the saxion dominates the Universe, the decay temperature cannot be
as high as $\GEV{12}$. Therefore the thermal production is not efficient in the saxion-dominated Universe. 
Although not pursued here, if the saxion does not dominate the Universe, the thermally
produced axions can explain the observed dark matter abundance if $T_R \sim \GEV{12-13}$, and also, 
they will contribute to warm dark matter. For lower $T_R$, the thermally produced axions contribute only a small fraction of 
the total dark matter density. 

While thermal production is negligibly small in our scenario, 
the axions can be copiously produced by coherent oscillations.
Neglecting the anharmonic effects~\cite{Turner:1985si,Lyth:1991ub,Visinelli:2009zm,Kobayashi:2013nva},
the axion abundance can be estimated as
\bea
\frac{\rho_a}{s} \;\simeq\; \frac{1}{8} T_R \lrfp{a_*}{M_p}{2},
\label{rho2s}
\eea
for the reheating temperature $T_R \lesssim \sqrt{m_a M_p} \sim 4 \times \GEV{6}$, where $a_*$ denotes the initial oscillation amplitude.
In this case the axion starts to oscillate before reheating.
The cosmic density is given by
\bea
\Omega_a h^2 \;\simeq\; 0.2 \lrf{T_R}{4{\rm\,GeV}} \lrfp{f_a}{5\times\GEV{14}}{2}  \lrfp{a_*/f_a}{0.2}{2},
\label{omegaa}
\eea
independent of the axion mass. For relatively low reheating temperature about GeV, the axion abundance falls in the right range
without fine-tuning of the initial misalignment angle $\theta_* \equiv a_*/f_a$.\footnote{
Strictly speaking, the decay constant for the axion potential could be slightly different from $f_a$, which is defined
by the coupling to the SM gauge sector (\ref{fa}). This however slightly modifies the required fine-tuning for
obtaining the right dark matter abundance, and our results are not changed.
} The amount of fine-tuning increases in proportion to $1/\sqrt{T_R}$.
On the other hand, for $T_R \gtrsim \sqrt{m_a M_p}$, the axion starts to oscillate after reheating, and
the abundance is approximately given by (\ref{rho2s}) and (\ref{omegaa}) with $T_R$ replaced with $\sqrt{m_a M_p}$.
For $T_R \gtrsim \GEV{6}$, the initial misalignment angle must be of order $10^{-4}$  for the right dark matter abundance.

As we shall see below, if the saxion dominates the Universe and decays into the SM sector,
the reheating temperature is determined by the saxion mass $m_\sigma$ and the decay constant $f_a$.
For instance,  $T_R \simeq 4$\,GeV is realized for $m_\sigma \simeq \GEV{6}$ and $f_a \simeq 5\times\GEV{14}$.

The axions produced by the initial misalignment mechanism are non-relativistic and therefore contribute to cold dark matter (CDM).
This should be contrasted to the sterile neutrinos with the same mass, which contribute to warm dark matter. 
Interestingly, as we shall see later in this section, the axions can be
also produced by the saxion decay, which may contribute to the HDM component. Therefore a mixed CDM$+$HDM model is possible
in our scenario.

\subsection{Saxion decay}

The saxion is stabilized by SUSY breaking effects, and its mass is of order the gravitino mass. 
If the inflation scale is larger than
or comparable to  the gravitino mass,  the position of the saxion during inflation is likely deviated from the low-energy minimum.
Then the saxion will start to oscillate with a large initial amplitude when the Hubble parameter becomes comparable to $m_\sigma$, and
may eventually dominate the Universe
after the inflaton decays.
For simplicity we assume that the Universe is dominated by the saxion before the axion commences its oscillations.

The saxion is coupled to the SM gauge sector through the interaction \REF{FWW}.
The relevant interactions are
\bea
{\cal L}_{\rm saxion} =
-\frac{g^2_a}{32\pi^2}
\frac{\sigma}{f_a} F^a_{\mu\nu}F^{a\mu\nu}
+
\left( \kappa \frac{g^2_a}{32\pi^2}
\frac{m_\sigma}{f_a}\,  \sigma \lambda_a\lambda_a + {\rm h.c.}\right),
\label{dilatonic}
\eea
with $g_a$ being the gauge coupling.
Here $\kappa$ is generally of order unity, and its precise value depends on the saxion stabilization
and the detailed structure of the K\"ahler potential.\footnote{
For instance, $\kappa = \sqrt{2}$ in the framework of Ref.~\cite{Higaki:2013lra}.
}
The typical gaugino mass is loop-suppressed compared to the gravitino mass in KKLT-type compactifications
with sequestered uplifting sector.
This is because the moduli have $F$-terms around $m_{3/2}/\ln(M_p/m_{3/2})$, making moduli mediation
comparable to   
anomaly mediation~\cite{Choi:2004sx,Endo:2005uy,Choi:2005uz}.\footnote{
It is possible to consider additional contributions to the gaugino masses
so that the saxion  decay into gauginos is kinematically forbidden.}
Therefore, the saxion decays into gauginos with a sizable branching
fraction, and it is not helicity suppressed~\cite{Endo:2006ix,moduli}.
The partial decay rates of the saxion into the SM gauge bosons and gauginos via (\ref{dilatonic}) are given by
\bea
\Gamma_{\sigma \to A_\mu A_\mu} &=& N_g \frac{\alpha^2}{256 \pi^3} \frac{m_\sigma^3}{f_a^2},
\\
\Gamma_{\sigma \to \lambda_a \lambda_a} &\simeq& N_g |\kappa|^2 \frac{\alpha^2}{256 \pi^3} 
\frac{m_\sigma^3}{f_a^2},
\eea
taking $g^2_a=4\pi\alpha$,
where $N_g = 12$ counts the SM gauge degrees of freedom, and we have omitted the phase space factor 
in the second equation.
In the following we will take $|\kappa| = 1$ as a reference value for simplicity.
The lightest SUSY particle (LSP) is overproduced in this case~\cite{Endo:2006ix}, as long as  the R-parity is conserved.
To avoid the overclosure of the Universe, we assume that the R-parity is broken. Alternatively the LSP abundance can be
suppressed in the presence of late-time entropy production, which is not pursued here.

The saxion also decays into a pair of axions with a rate
\bea
\Gamma_{\sigma \rightarrow aa} = 
\frac{1}{64\pi}
\frac{\langle\partial^3_\Phi K \rangle^2}{\langle \partial^2_\Phi K\rangle^3}
m^3_\sigma,
\label{axdecay}
\eea
which can be comparable to the decay rate into the SM gauge sector. To see this let us calculate 
the ratio of the rates,
\bea
\frac{\Gamma_{\sigma \rightarrow aa}}{\Gamma_{\sigma \to A_\mu A_\mu}+\Gamma_{\sigma \to \lambda_a \lambda_a} }
&\simeq& 0.33   \lrf{2}{k^2(1+|\kappa|^2)} \lrf{12}{N_g} \lrfp{1/25}{\alpha}{2}
\left(\frac{\langle \partial^3_\Phi K \rangle}{\langle \partial^2_\Phi K\rangle}\right)^2.
\eea
Thus produced axions lead to cosmological problems, which is a general feature of such moduli fields stabilized
by SUSY breaking effects: the so-called ``the moduli-induced axion problem"~\cite{Higaki:2013lra}.
Those axions are ultra-relativistic at the production, and lose the kinetic
energy as the Universe expands, and eventually become non-relativistic as they have a non-zero mass about $7$\,keV.
They are subject to the BBN constraint on the additional effective neutrino species $\Delta N_{\rm eff}$~\cite{Steigman:2012ve}
as well as the HDM constraint set by the large-scale structure
observation~\cite{Wyman:2013lza,Hamann:2013iba, Battye:2013xqa}.
The axion contribution to $\Delta N_{\rm eff}$ can be suppressed if there is an approximate $Z_2$ symmetry under which $\Phi$ changes the sign
in the underlying theory. For the moment we set 
$\langle \partial^3_\Phi K \rangle=0$ for simplicity. 
We will return to the case of $\langle \partial^3_\Phi K \rangle\neq 0$
when we discuss the HDM constraint on the axions produced by the saxion decay.

On the other hand, the saxion decay into a pair of gravitinos or axinos can be kinematically forbidden as
these particles have a comparable mass. Therefore the notorious moduli-induced
gravitino problem~\cite{moduli,Dine:2006ii,Endo:2006tf} can be avoided in our scenario.
This is indeed the case in the moduli stabilization discussed in Sec.~\ref{sec:2}.

Assuming that the saxion mainly decays into the SM sector via the dilatonic coupling, the decay temperature 
is estimated as
\bea
T_R &\simeq & 4 {\rm\,GeV}\, \lrfp{g_*(T_R)}{106.75}{-\frac{1}{4}}  
\lrfp{m_\sigma}{\GEV{6}}{\frac{3}{2}} \lrfp{f_a}{5\times \GEV{14}}{-1},
\eea
where $g_*(T_R)$ counts the relativistic degrees of freedom in the plasma at the saxion decay.
Combined with \REF{omegaa}, one can see that the right amount of axion dark matter is produced for the saxion
mass about $\GEV{6}$, the decay constant $f_a \simeq 5\times \GEV{14}$ and the initial misalignment 
angle $\theta_* \sim 0.2$.
For a heavier mass of the saxion, $\theta_*$ should be suppressed in proportion to $m_\sigma^{-3/4}$.

\subsection{Hot dark matter constraint}

The axions produced by the saxion decay may contribute to the HDM component.
This issue was discussed in detail in Ref.~\cite{Jeong:2013oza}, motivated by  the cosmological
preference for a HDM component~\cite{Wyman:2013lza,Hamann:2013iba, Battye:2013xqa}.

The properties of HDM can be characterized by the abundance and the effective mass.
The abundance is often expressed in terms of the additional neutrino species, $\Delta N_{\rm eff}$,
defined by  the ratio of the HDM energy density to the energy density of single neutrino species in the relativistic limit.
The contribution of axions to $\Delta N_{\rm eff}$ is given by~\cite{Choi:1996vz,Jeong:2012np}
\bea
\Delta N_{\rm eff}
&=&  \frac{43}{7} \lrfp{g_{* \nu}}{g_*(T_R)}{\frac{1}{3}} \frac{B_a}{1-B_a},
\eea
where $B_a$ denotes the branching fraction into axions, and
$g_{* \nu} = 10.75$. For instance, $\Delta N_{\rm eff} = 0.6$ is obtained for
$B_a \simeq 0.17$ and $g_*(T_R)=106.75$. Note that the abundance is fixed by
 $\langle\partial^3_\Phi K \rangle^2/\langle \partial^2_\Phi K\rangle^3$, independent of the saxion mass.
 In general, $\Delta N_{\rm eff} = {\cal O}(0.1-1)$ is expected~\cite{Higaki:2013lra}.

The timing when the axions become non-relativistic can be estimated by the effective
hot dark matter mass~\cite{Jeong:2013oza},
\bea
m_a^{\rm (eff)} &=& \frac{\pi^4 }{30 \zeta(3)} \Delta N_{\rm eff}
\frac{T_R}{m_\sigma/2}
 \lrfp{g_*(T_R)}{g_{*\nu}}{\frac{1}{3}} m_a,
\eea
which roughly coincides with a mass of thermally produced HDM with the abundance $\Delta N_{\rm eff}$.
Namely, the axion HDM becomes non-relativistic when the cosmic temperature is comparable to $m_a^{\rm (eff)}$.
As the axions are ultra-relativistic at the production, they behave like HDM with an effective mass much lighter
than their actual mass.
For the parameters of our interest, it is given by
\bea
m_a^{\rm (eff)} &\simeq&  0.2{\rm \,eV}
 \lrf{\Delta N_{\rm eff}}{0.6} \lrf{m_a}{7\,{\rm keV}}
 \lrfp{m_\sigma}{\GEV{6}}{\frac{1}{2}}
\lrfp{f_a}{5\times \GEV{14}}{-1},
\eea
where we have set $g_*(T_R) = 106.75$ and $\alpha=1/25$.

It is interesting to compare the above values of $\Delta N_{\rm eff}$ and $m_a^{\rm (eff)}$
with the recent results of Refs.~\cite{Wyman:2013lza,Hamann:2013iba, Battye:2013xqa}.
According to Ref.~\cite{Hamann:2013iba}, a combination
of Planck data, WMAP-9
polarization data, measurements of the BAO scale, the HST measurement of the $H_0$, Planck galaxy
cluster counts and galaxy shear data from the CFHTLens survey yields
\bea
\label{obsnf}
\Delta N_{\rm eff} &=& 0.61 \pm 0.30,\\
\label{obsmf}
m_{\rm HDM}&=& (0.41 \pm 0.13)\, {\rm eV},
\eea
at $1\sigma$. Note however that, precisely speaking, we cannot directly apply the observational results
 (\ref{obsnf}) and (\ref{obsmf}) to the case of the axion HDM, due to the different momentum distribution
 as well as the numerical coefficient in the definition of the effective mass. Nevertheless it is intriguing that
 our set-up can naturally implement the HDM, which seems favored by the observations.

 If the preference for  a HDM component is simply an artifact of the systematic
 uncertainties of various observations,  the axion HDM abundance must be sufficiently small.
 This can be realized  by suppressing $\langle \partial^3_\Phi K\rangle$ without severe fine-tuning.
For instance, 
$\langle \partial^3_\Phi K \rangle/\langle \partial^2_\Phi K\rangle \sim 0.1$
would give $\Delta N_{\rm eff} \sim 0.01$,
which has only negligible impact on the large-scale structure.

\subsection{Isocurvature constraints}
The axion acquires  quantum fluctuations during inflation, giving rise to the CDM isocurvature perturbations,
as in the case of the QCD axion.
The mixture of the CDM isocurvature perturbations is tightly constrained by the CMB observations~\cite{Ade:2013uln} as
\bea
\frac{{\cal P}_S}{{\cal P}_{\cal R}+{\cal P}_S}  &<& 0.039~~~(95\%{\rm \,CL},~~Planck + {\rm WP})
\eea
where ${\cal P}_S$ and ${\cal P}_{\cal R} $ are the power spectrum for the
isocurvature and curvature perturbations, respectively. The Planck normalization reads ${\cal P_R} \simeq 2.2 \times 10^{-9}$.

\begin{figure}[t]
\begin{center}
\begin{minipage}{16.4cm}
\centerline{
{\hspace*{0cm}\epsfig{figure=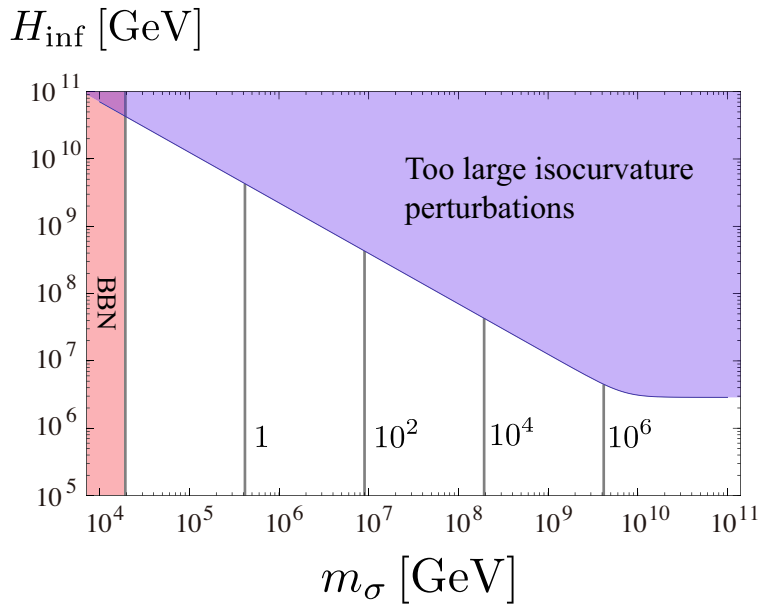,angle=0,width=10cm}}
}
\caption{
The shaded regions are excluded by too large isocurvature perturbations of the axion dark matter (upper right triangle region) or by
too low reheating temperature, $T_R \lesssim 10$\,MeV, which would spoil the BBN (left rectangular region). Here we take $r=1$, namely, the axion
explains all the dark matter, and $f_a = 5\times \GEV{14}$. The contours for
the reheating temperature, $T_R = 1, 10^2, 10^4, 10^6$ GeV, are also shown. The isocurvature constraint becomes insensitive to
$m_\sigma$ for $T_R \gtrsim \sqrt{m_\sigma M_p}$.
}
\label{fig:iso}
\end{minipage}
\end{center}
\end{figure}

In our axion dark matter model, the power spectrum of the isocurvature
perturbations is estimated by~\cite{Kobayashi:2013nva}\footnote{
The saxion is considered to be deviated from the low-energy minimum during inflation,
which may change the isocurvature perturbations  by a factor of ${\cal O}(1)$.
}
\bea
\label{num0}
{\cal P}_S &=& \left(r \dd{\ln \Omega_a}{\theta_*} \frac{H_{\rm inf}}{2\pi f_a}\right)^2,
\label{num}
\eea
where $r$ denotes the fraction of the axion density to the total dark matter density, $\theta_* \equiv a_*/f_a$ represents
the initial misalignment angle, and $H_{\rm inf}$ is the Hubble parameter during inflation.
Assuming the axion explains the total dark matter density, i.e., $r=1$, and $\Omega_a \propto \theta_*^2$ as in Eq.~(\ref{omegaa}),
we obtain  ${\cal P}_S \simeq (H_{\rm inf}/\pi a_*)^2$. Then the observational bound reads
\bea
H_{\rm inf}&\lesssim& 3\times \GEV{9} \lrf{a_*/f_a}{0.2} \lrf{f_a}{5\times \GEV{14}}.
\eea

In Fig.~\ref{fig:iso} we show the region excluded by the isocurvature constraints in the plane of the saxion mass and the Hubble
parameter during inflation.  Among  the 5 parameters, $H_{\rm inf}$, $m_\sigma$, $m_a$, $f_a$, and $\theta_*$,
the dark matter abundance and  the observed X-ray line fix $3$ of them, the axion mass, the decay constant,
and a combination of $m_\sigma$ and $\theta_*$.  Then we can express the initial misalignment angle and the reheating
temperature as a function of the saxion mass by using (\ref{omegaa}), as one can see
the contours of $T_R = 1, 10^2, 10^4, 10^6$\,GeV in the figure. We have set  $f_a = 5 \times \GEV{14}$.
We also show the region excluded by the  big bang nucleosynthesis where the reheating temperature is
below $10$\,MeV~\cite{Kawasaki:1999na}.

The isocurvature perturbations  can be suppressed if the axion acquires a
large mass during inflation by some non-perturbative dynamics, which disappears after inflation.
For instance, if the Higgs field has a large expectation value during inflation, the QCD interactions become strong at an intermediate or 
high energy scale, generating a heavy mass to the axion~\cite{Jeong:2013xta}.

\subsection{Baryogenesis}
In our scenario the saxion dominates the Universe and reheats the SM sector. Because of a relatively large decay constant,
the reheating temperature tends to be low, as one can see from Fig.~\ref{fig:iso}. This could be an obstacle for creating
the right amount of baryon asymmetry. Here we briefly mention a couple of possible ways to generate baryon asymmetry.

The saxion decays into the SUSY particles with an unsuppressed rate, and so, the LSPs would overclose the Universe
unless the R-parity is broken. To avoid this problem  we have assumed that the R-parity is explicitly broken.
In fact,  the right amount of baryon asymmetry can be generated through CP violating decay of gluino into quark and
squark followed by baryon-number violating squark decay~\cite{Ishiwata:2013waa}.
(See also Refs.~\cite{Cline:1990bw,Mollerach:1991mu}.)
For this saxion-induced baryogenesis to work, we introduce the R-parity and baryon-number violating operator,
\beq
W = \frac{1}{2} \lambda_{ijk} U^c_i D^c_j D^c_k,
\eeq
where $U^c_i$ and $D^c_j$ are the SU(2)$_L$ singlet up-type and
down-type quarks, respectively, and $i,j,k$ are flavor indices.
The required  CP phase  between the gaugino mass and the A-term of the above operator can be
generated from the relative phase between the non-perturbative terms\footnote{ In \EQ{Wxi}, we can
add exponential terms of $X_i$ without modifying the discussion so far. Then the relative phases among
the non-perturbative terms source the CP phase.
}
through a mixed modulus-anomaly mediation of the heavy moduli $X_i$~\cite{Ishiwata:2013waa}.
The resultant baryon asymmetry is given by
\begin{eqnarray}
\frac{n_B}{s} &\simeq&
3\times 10^{-10} \frac{|\kappa|^2}{\sqrt{1+|\kappa|^2}} 
\left(\frac{m_\sigma}{10^6~{\rm GeV}}\right)^{1/2}
\lrfp{f_a}{5 \times \GEV{14}}{-1}
\left(\frac{\epsilon_B}{10^{-4}}\right),
\end{eqnarray}
where we have set $g_*(T_R) = 106.75$, and $\epsilon_B$ denotes the effective baryon number generated by a single gluino decay.
Also we assumed that only $\lambda_{332}$ is non-zero and of order unity, and in this case, the efficiency coefficient 
is given by $\epsilon_B \lesssim 10^{-2}$, where the upper bound is saturated for the maximal CP phase. 
Therefore,  the right amount of baryon asymmetry can be generated for the
saxion mass of our interest. For the saxion mass of ${\cal O}(10^{4-5})$\,GeV, the typical soft mass for the SUSY SM particles
is in the TeV range. Then, some of them may be within the reach of LHC, and also, a part of the parameter space can be probed
by the dinucleon decay search experiment and the measurement of the electric dipole moments of neutron and electron~\cite{Ishiwata:2013waa}.

For the saxion mass $m_\sigma \gtrsim \GEV{10}$, the reheating temperature becomes high enough for non-thermal
leptogenesis~\cite{Fukugita:1986hr} to work, if the saxion mainly decays into the right-handed neutrinos~\cite{Jeong:2013axf}.
Another possibility is to generate a  large amount of the baryon asymmetry by the Affleck-Dine
mechanism~\cite{Affleck:1984fy,Dine:1995kz}.

\section{Discussion and Conclusions}
There appear many moduli fields in the low energy through compactifications of extra dimensions in superstring theories.
Some of the moduli fields may remain light after the closed string flux is turned on.
We have focused on a modulus field which is stabilized by the SUSY breaking effect and its axion component remains
much lighter than the saxion component. As long as the strong CP problem is solved by the string-theoretic QCD axion,
there must be at least one such modulus field, and  in general, there might be more. As such moduli fields
tend to be lighter than those stabilized by the non-perturbative effects {\it a la} KKLT, they likely play an important cosmological role.

We have proposed a scenario in which the saxion component of such modulus field
dominates the energy density of the Universe and reheats
the SM sector via its dilatonic couplings, while its axion partner contributes to dark
matter decaying into photons via the same dilatonic coupling to photons in SUSY.  
The point is that both the reheating of the Universe and the decay of dark matter into photons are induced by 
the same supermultiplet (i.e. saxion and axion) through the same operator in SUSY. 
This observation partially explains why dark matter decays into photons at all. If there are light axions, one of them can easily
explain the dark matter abundance as  the axions are copiously generated by coherent oscillations.
Then, there is no special reason why the axion dark matter should be coupled to photons. The situation changes if the bosonic partner,
the saxion, dominates the Universe and reheats the SM sector through the same operator in SUSY.
In this case, the axion dark matter must be coupled to the SM sector, in order for successful reheating.
In other words, the decaying dark matter can be a probe of the reheating of the Universe.

We have also discussed the saxion decay process,
the HDM constraint on the axions produced by the saxion decay, the isocurvature constraint on the axions produced by coherent
oscillations, and the baryogenesis scenarios. Some of our results, especially those about the nature of axion dark matter (i.e. abundance, lifetime and isocurvature constraints), can be straightforwardly applied to the case in which the saxion does not dominate the Universe.
This is likely the case e.g. if the Hubble parameter during inflation is smaller than the saxion mass.

Interestingly, for the axion mass $m_a \simeq 7$\,keV
and the decay constant $f_a \simeq 10^{14-15}$\,GeV,   the recently discovered X-ray line at $3.5$\,keV in the XMM Newton
X-ray observatory data  can be explained by the decay of the axion dark matter. The suggested value of the decay constant is within the expected range for the string-theoretic axion. It is of course possible to consider field-theoretic axions or pseudo Nambu-Goldstone bosons of mass $7$\, keV  which have couplings to photons
with a similar strength.
The detailed  X-ray line search in future may
 not only probe the nature of dark matter but also unravel the very early history of our Universe as well as physics close to the
 GUT scale.

\section*{Acknowledgment}
This work was supported by Grant-in-Aid for  Scientific Research on Innovative
Areas (No.24111702, No. 21111006, and No.23104008) [FT], Scientific Research (A)
(No. 22244030 and No.21244033) [FT], and JSPS Grant-in-Aid for Young Scientists (B)
(No. 24740135 [FT] and No. 25800169 [TH]), and Inoue Foundation for Science [FT].
This work was also supported by World Premier International Center Initiative
(WPI Program), MEXT, Japan [FT].

\end{document}